\documentstyle[11pt,a4]{article}
\begin{document}
\begin{center}
{\bf BEYOND NONLINEAR SCHR\"ODINGER EQUATION APPROXIMATION FOR AN 
ANHARMONIC CHAIN WITH HARMONIC LONG RANGE INTERACTIONS}
\end{center}
\vskip .3cm
\begin{center}
{\bf D. Grecu,}\footnote{E-mail 
address: ~~~dgrecu@theor1.theory.nipne.ro} {\bf Anca Vi\c sinescu, 
A.S. C\^arstea}\\
Department of Theoretical Physics\\
National Institute of Physics and Nuclear Engineering\\
Bucharest, Romania
\end{center}
\vskip .3cm

\begin{abstract}
Multi scales method is used to analyze a nonlinear 
differential-difference equation. In the order $\epsilon^{3}$ the NLS 
eq. is found to determine the space-time evolution of the leading 
amplitude. In the next order this has to satisfy a complex mKdV eq. 
(the next in the NLS hierarchy) in order to eliminate secular terms. The 
zero dispersion point case is also analyzed and the relevant equation 
is a modified NLS eq. with a third order derivative term included.
\end{abstract}

Many one-dimensional systems of biological interest are very 
complicated structures, formed from complexes of atoms - we shall call 
them "molecules" - connected by  hydrogen bounds. It is usually assumed 
that only one of the intra-molecular excitations plays an active role in 
the storage and transport of e\-ner\-gy in these systems. In the case of 
$\alpha$-helix structure in protein this corresponds to the amide I 
vibration (C=O stretching). We shall call this intra-molecular 
excitation the vibronic field. Localized excitations of solitonic type 
can exist in these systems, due to a nonlinear interaction between the 
vibronic field and the acoustic phonon field, describing the molecule 
oscillations along the chain. The simplest model starts from a 
Fr\"olich Hamiltonian and with an ansatz - coherent state approximation 
- for the state vector describing this type of localized excitation 
(\cite{1} - \cite{3} and references therein).

After eliminating phonon variables a nonlinear differential (time) - 
difference (space) equation for the vibronic coordinate is obtained;
\begin{equation}
L(\{y_{n}\})=G(\{y_{n}\})
\end{equation}
where $L$ is the linear part and $G$ the nonlinear one. For the 
specific example we have in mind, originating from Takeno's model 
\cite{2}, $L$ and $G$ are given by
\begin{equation}
L(\{y_{n}\})={d^{2}y_{n}\over dt^{2}} +\omega_{0}^{2}y_{n}-\sum_{m\ne 
n} J_{mn}y_{m}
\end{equation}
\begin{equation}
G(\{y_{n}\})={1\over 2}Ay_{n}(y^{2}_{n+1}+y^{2}_{n-1}) - By^{3}_{n}.
\end{equation}
In the linear part the last term is a long range interaction between 
vibrons, and we shall assume that $J_{mn}$ decreases exponentially 
(Kac-Baker model)
\begin{equation}
J_{mn}=J_{\vert m-n \vert}=\omega^{2}_{LR}{1-r\over 2r} e^{-\gamma\vert 
m-n \vert}, ~~~~~~~r=e^{-\gamma}.
\end{equation}
The first term in r.h.s. of $G$ results from the nonlinear interaction 
between vibrons and phonons, while the second one from a quartic 
anharmonicity in the vibron Hamiltonian.

The linear equation admit plane wave solutions $ e^{i\theta} $, $ \theta = 
kan-\omega t $, where $\omega (k)$ is given by the dispersion relation
\begin{eqnarray}
&&D(\omega k)=\omega^{2}(k)-\left(\omega_{0}^{2}-2\sum^{\infty}_{p=1}
J_{p} \cos kap\right)= \nonumber \\
&&\omega^{2}-\left(
\omega^{2}_{0}-\omega^{2}_{LR}{1-r\over 2r}\left({\sinh \gamma \over 
\cosh \gamma - \cos ka}-1\right)\right)
\end{eqnarray}
describing an optical vibrational branch with $\omega^{2}(k)$ a 
monotonously increasing function of $k$ from
$\omega^{2}(0)=\omega^{2}_{0}-\omega^{2}_{LR}$ to 
$\omega^{2}(\pi)=\omega_{0}^{2}+\omega^{2}_{LR}{1-r\over 1+r}$. We 
shall assume that a no-resonance condition takes place
\begin{equation}
D_{\nu}=D(\nu \omega,\nu k) \ne 0, ~~~~\nu\in N^{*}, ~~\nu\ne 1.
\end{equation}

It is well known that the effect of a weak nonlinearity occurs at large 
space-time scales, determining a redistribution of energy on higher 
harmonics, and a modulation of amplitude. In order to investigate these 
effects we shall use the multi-scales method (reductive perturbation 
method) \cite{4}. The method starts by introducing slow space-time 
variables
\begin{equation}
x=\epsilon an, ~~~~~~~~~~~~~~~t_{j}=\epsilon^{j}t
\end{equation}
and expanding $y_{n}$ in an asymptotic perturbative series, which due 
to the form (3) of the nonlinearity $G$ is given by 
\begin{equation}
y_{n}=\sum^{odd}_{\nu=1}e^{i\nu\theta}\sum^{\infty}_{p=\nu}\epsilon^{\nu}
Y_{p,\nu}(x;t_{1},t_{2},...)+c.c
\end{equation}

Recently several papers have used this method to discuss the 
propagation of quasi-monochromatic waves in weakly nonlinear media 
\cite{5}-\cite{7}, or of long surface waves in shallow waters \cite{8}. 
Very interesting are the conclusions concerning the role played by the 
NLS hierarchy \cite{7}, or the KdV hierarchy \cite{8} in eliminating 
the secular terms which would destroy the asymptotic character of the 
perturbative series. Of special interest for the present paper is the 
reference \cite{7}, which will be followed as close as possible.

In calculating the time derivative we have to take into account that $t$ 
appears in $\theta$ as well as in the slow time variables $t_{1}, 
t_{2},$... Also in writing the expressions for $y_{n\pm 1}, y_{m}$ we 
have to expand the corresponding amplitudes around the point $n$. Taken
these precautions  the calculations are straightforward (although 
quite tedious in the higher orders): the asymptotic expansion is 
introduced in (1), (2), (3) and the coefficient of each power of 
$\epsilon $ and each harmonic $e^{i\nu\theta}$ is equated with zero.

In the first order in $\epsilon $ we re-obtain the dispersion relation 
(5). In the order $\epsilon^{2}$ the amplitude $Y_{1,1}$ has to satisfy 
the equation
\begin{equation}
L_{+}Y_{1,1}=\left({\partial \over\partial t_{1}}+ v_{g}{\partial\over 
\partial x}\right) Y_{1,1}=0
\end{equation}
and consequently $Y_{1,1}$ will depend only on the variable $\xi=x-vt$, 
where $v_{g}={d\omega\over dk}=\omega_{1}$ is the group velocity.

In the next order $\epsilon^{3}$, from the terms proportional with 
$e^{i\theta}$ we get
\begin{equation}
L_{+}Y_{2,1}={\partial Y_{1,1}\over\partial t_{2}}-K_{2}(Y_{1,1})
\end{equation}
\begin{equation}
K_{2}(Y_{1,1})=i\omega_{2}\left({\partial^{2}Y_{1,1}\over \partial 
\xi^{2}}+q\vert Y_{1,1} \vert^{2}Y_{1,1}\right)
\end{equation}
Here $\omega_{n}={1\over n!}{d^{n}\omega \over dk^{n}}$ and $q={A\over 
2\omega}(2+\cos 2ka - 3{B\over A})$. As the r.h.s. of (10) is in the 
null space of $L_{+}, Y_{2,1}$ will blow up linearly in $t_{1}$ unless 
the r.h.s. is strictly equal with zero, i.e. $Y_{1,1}$ has to evolve in 
$t_{2}$ according to the cubic nonlinear Schr\"odinger equation 
$(c={q\over 2\omega_{2}})$
\begin{equation}
{\partial Y_{1,1}\over \partial t_{2}}=i\omega_{2}\left({\partial^{2}Y_{1,1}
\over \partial \xi^{2}}+2c\vert Y_{1,1} \vert^{2}Y_{1,1}\right)
\end{equation}
In this case $Y_{2,1}$ will depend also on the characteristic 
coordinate $\xi$ only. From terms proportional with the third harmonic 
$e^{3i\theta}$ one obtains
\begin{equation}
D_{3}Y_{3,3}+(A\cos 3ka - B)Y^{3}_{1,1}=0
\end{equation}
and due to the no-resonance condition (6) it is an algebraic equation 
giving $Y_{3,3}$ in terms of $Y_{1,1}$. The same thing happens for all 
the higher harmonics and the corresponding amplitudes $Y_{p,\nu}$ can 
be explicitly written in terms of {$Y_{p,1}$} and their derivatives. 
Therefore we shall concentrate our attention to the amplitudes 
$Y_{p,1},$ related to the first harmonic $e^{i\theta}$.

The solution of the NLS eq. (12) depends on the sign of $\omega_{2}$ 
and $q$. As $\omega_{1}$ vanishes at $k=0$ and $k=\pi$, there is a 
point $k_{c}\in (0,\pi)$ for which $\omega_{2}(k_{c})=0$. If $k<k_{c}
(k>k_{c})$ we have $\omega_{2}>0 (\omega_{2}<0)$. The sign of $q$ 
depends on the constants $A$ and $B$. For $A>0$ and $B<{A\over 3}$ it 
is always positive, while for $B>A$ it is negative. Depending on the 
sign of $\omega_{2}$ and $q$ the NLS eq. (12) can have bright or dark 
soliton solutions.

In the order $\epsilon^{4}$ from the terms proportional with 
$e^{i\theta}$ we get
\begin{equation}
{\partial Y_{2,1}\over \partial t_{2}}-K_{2}^{\prime}(Y_{2,1})=
-{\partial Y_{1,1}\over \partial 
t_{3}}+\omega_{3}{\partial^{3}Y_{1,1}\over \partial 
\xi^{3}}-2c{\omega_{1}\omega_{2}\over \omega}Y_{1,1}{\partial\vert 
Y_{1,1}\vert^{2}\over \partial\xi}+q_{1}\vert Y_{1,1}\vert^{2}{\partial 
Y_{1,1}\over \partial\xi}.
\end{equation}
Here
\begin{equation}
K^{\prime}_{2}(Y_{1,1}) = i\omega\left({\partial^{2}Y_{2,1}\over 
\partial\xi^{2}}+2c(Y_{1,1}^{2}Y^{*}_{2,1}+2\vert 
Y_{1,1}\vert^{2}Y_{2,1})\right)
\end{equation}
is the Frechet derivative of $K_{2}$, and $q_{1}={dq\over dk}$.
The l.h.s. of (14) is the linearized NLS eq. It is well known that the 
commuting symmetries {$\sigma_{j}$} of the NLS eq. are solutions of 
this equation. As they are important for our further discussion we 
remained the expression of the first ones (by $\Psi$ we shall denote a 
solution of the NLS eq.)
\begin{eqnarray}
&& \sigma_{0}=-i\Psi,~~~~~~~~~~~~~~~~~~~~~~~~\sigma_{1}={\partial 
\Psi\over \partial\xi} \nonumber \\
&& \sigma_{2}=i({\partial^{2}\Psi\over \partial\xi^{2}}+2c\vert 
\Psi\vert^{2}\Psi) ~~~~~~~~\sigma_{3}=-({\partial^{3}\Psi\over 
\partial\xi^{3}}+6c\vert \Psi\vert^{2}{\partial \Psi\over \partial \xi})
\end{eqnarray}
The eq. (14) is a forced linear equation for $Y_{2,1}$. It is necessary 
to identify secular terms in the r.h.s. of (14) and then to fix the 
$t_{3}$ dependence of $Y_{1,1}$ in such a way to eliminate their 
effect. These secular terms have to be found between the members of the 
null space of linearized NLS eq., i.e. between the commuting 
symmetries {$\sigma_{j}$}. Indeed if such a symmetry $\sigma$ would 
exists it will generate a $t_{2}\sigma$ contribution to $Y_{2,1}$, and 
the asymptotic character of the expansion (8) would be destroyed in a 
time $t_{2}=O(\epsilon^{-1})$. Two such symmetries $(\sigma_{0}, 
\sigma_{3})$ are easily seen in the r.h.s. of (14), if it is written 
in the form
\begin{equation}
-{\partial Y_{1,1}\over \partial 
t_{3}}+\omega_{3}\left({\partial^{3}Y_{1,1}\over 
\partial\xi^{3}}+6c\vert Y_{1,1}\vert^{2}Y_{1,1}\right) +N(Y_{1,1})
\end{equation}
where
$$ N(Y_{1,1})=-2c{\omega_{1}\omega_{2}\over \omega}Y_{1,1}{\partial 
\vert Y_{1,1}\vert^{2}\over \partial\xi}+(q_{1}-6c\omega_{3})\vert 
\Psi_{1,1}\vert^{2}{\partial \Psi_{1,1}\over \partial \xi}. $$
In order to avoid this secular behaviour we require that the $t_{3}$ 
dependence of $Y_{1,1}$ is given by the following complex modified KdV 
equation
\begin{equation}
-{\partial Y_{1,1}\over \partial t_{3}}+\omega_{3}\left({\partial^{3}Y_{1,1}
\over \partial \xi^{3}}+6c\vert Y_{1,1} \vert^{2}Y_{1,1}\right)=0
\end{equation}
which is the next equation in the NLS hierarchy. The influence of the 
rest $N(Y_{1,1})$ on $Y_{2,1}$ can be further treated using a Green 
function formalism \cite{9}.

Let us consider a single soliton solution \cite{10}
\begin{equation}
\Psi = 2{P_{1}\over \sqrt{c}}{e^{-i\phi}\over \cosh z}
\end{equation}
$$ \phi(\xi,t_{2})=2S_{1}\xi+4\omega_{2}(S^{2}_{1}-P_{1}^{2})t_{2}+\phi_{
0} $$
$$ z(\xi,t_{2})=2P_{1}(\xi -\xi_{0}+4\omega_{2}S_{1}t_{2}) $$
where $S_{1}, P_{1}$ are the real and imaginary part of the complex 
eigenvalue $ \zeta_{1}=S_{1}+iP_{1}$ in the inverse scattering 
transform method, and $\phi_{0}, \xi_{0}$ are the initial phase and the 
initial position of the soliton. Applying the above procedure, in order 
to eliminate the possible secularities, the soliton parameters must be 
$t_{3}$-dependent. This dependence can be found introducing (19) in 
(18). The complex eigenvalue will remain unchanged, while for 
$\phi_{0}, \xi_{0}$ the following linear equations are found \cite{7}
\begin{eqnarray}
&& {d\phi_{0}\over dt_{3}}=-8\omega_{3}S_{1}(S^{2}_{1}-3P^{2}_{1}) 
\nonumber \\
&& {d\xi_{0}\over dt_{3}}=-8\omega_{3}P_{1}(3S^{2}_{1}-P^{2}_{1})
\end{eqnarray}
A similar analysis was given by Kodama \cite{11}, and the same results 
are obtained using the direct perturbation method of Keener and 
McLaughlin \cite{9}. More complex situations and details will be 
published elsewhere.

Let us consider now the situation when $\omega_{2}=0$, i.e. the 
propagation of a wave with the wave vector $k_{c}$. As $\omega_{1}$ has 
a maximum at this point it represents the wave propagating with the 
highest group velocity. A similar situation is encountered in the case 
of pulse propagation in nonlinear optical fibers where this point is 
known as the "zero dispersion point" (ZDP) \cite{12}-\cite{14}. The 
power required to generate an optical soliton is minimal in this point, 
and its evolution in space and time is governed by a modified NLS eq., 
with a third order derivative included. We shall show that a similar 
situation appears in the present case.

In applying the multiple scale method we shall use the same asymptotic 
expansion (8) for the vibronic variable. Then the nonlinearity 
contribution begins with terms of order $\epsilon^{3}$. To have 
contributions of the same order from third order derivatives we have to 
change the scaling of the $\xi$ variable, namely
\begin{equation}
\xi=\epsilon^{{2\over 3}}(an-\omega_{1}t).
\end{equation}
We have to take into account also a dependence of the phase $\theta$ of 
the propagating wave on the slow variable $\xi$. Defining the local 
wave number $k$ as the derivative of the phase $\theta$ with respect to 
$(an)$ we find that $k$ is slightly different from $k_{c}$, and the 
simplest choice is
\begin{equation}
k=k_{c}(1+\epsilon^{{2\over 3}}).
\end{equation}
Expanding all the quantities depending on $k$ around the point $k_{c}$ 
in the order $\epsilon^{3}$ the following equation is found for the 
leading amplitude $Y_{1,1}\rightarrow \Psi,$
\begin{equation}
i\Psi_{T} +3\Psi_{XX}+i\Psi_{XXX}+Q\vert \Psi \vert^{2}\Psi = 0
\end{equation}
where $X=k_{c}\xi,~~ T=\Omega t_{2}, ~~\Omega =\omega_{3}k^{3}_{c}$ and
$Q={q\over \Omega}.$
It has the same form as the equation describing the propagation of 
nonlinear pulses in optical fibers in the ZDP region 
\cite{12}-\cite{14}. In our case it makes the transition between the 
two regions, where bright and dark solitons exist. It seems that it is 
not completely integrable, but some analytical and numerical results 
suggest that some long-living localized excitations exist \cite{14}. 
Further investigations are necessary.
\vskip .3cm
{\it Acknowledgements: Two of the authors (DG and ASC) would like to 
thank the Organizing Committee of NEEDS 99 for financial support. Helpful 
discussions with Professors A. Degasperis, L. Kalyakin, V.V. Konotop, 
P.M. Santini are kindly acknowledged.}


\begin{thebibliography}{99}
\bibitem{1} A.S. Davydov, "{\it Solitons in Molecular Systems}", 
(Reidel, Dordrecht, 1985).
\bibitem{2} S. Takeno, Prog. Theor. Phys., {\bf 73}, 853 (1985); {\bf 
75}, 1 (1986); 

J. Phys. Soc. Japan {\bf 59}, 3127 (1990).
\bibitem{3} P.L. Christiansen, A.C. Scott (eds) - "{\it Davydov's 
Soliton Revisited. Self Trapping of Vibrational Energy in Protein}" 
(NATO ASI Ser. B {\bf 243}, Plenum Press, NY, 1990).
\bibitem{4} T. Taniuti, Suppl. Prog. Theor. Phys. {\bf 55}, 1 (1974).
\bibitem{5} M. Remoissenet, N. Flytzanis, J. Phys. C: Solid State Phys. 
{\bf 18}, 1573 (1985);

M. Remoissenet, Phys. Rev. B {\bf 33}, 2386 (1986).
\bibitem{6} V.V. Konotop, Phys. Rev. E {\bf 53}, 2843 (1996).
\bibitem{7} A. Degasperis, S.V. Manakov, P.M. Santini, Physica {\bf D 
100}, 187 (1997).
\bibitem{8} R.A. Kraenkel, M.A. Manna, J.G. Pereira, J. Math. Phys. 
{\bf 36}, 307 (1995).
\bibitem{9} J.K. Keener, D.W. McLaughlin, Phys. Rev. A {\bf 10}, 777 
(1977);

J. Math. Phys. {\bf 18}, 2008 (1977).
\bibitem{10} M.J. Ablowitz, H. Segur, "{\it Solitons and Inverse 
Scattering Transform}" (SIAM, Philadelphia, 1981).
\bibitem{11} Y. Kodama, J. Phys. Soc. Japan {\bf 45}, 311 (1978).
\bibitem{12} P.K.A. Wai, C.R. Menyuk, Y.C. Lee, H.H. Chen, Optics Lett. 
{\bf 11}, 464 (1986);

P.K.A. Wai, H.H. Chen, Y.C. Lee, Phys. Rev. A {\bf 41}, 426 (1990).
\bibitem{13} M. Klauder, E.W. Laedke, K.H. Spatschek, S.K. Twitsyn, 
Phys. Rev. E {\bf 47}, R 3844 (1993).
\bibitem{14} I.M. Uzunov, M. G\"olles, F. Lederer, Phys. Rev. E {\bf 
52}, 1059 (1995).

\end{thebibliography}
\end{document}